\documentclass[a4paper,11pt]{article}
\pdfoutput=1 

\usepackage{jcappub} 

\usepackage[T1]{fontenc} 

\title{\boldmath String length constraining from the stochastic gravitational waves background}

\author[a]{Hongguang Zhang,}
\author[a]{Xilong Fan,\footnote{Corresponding author.}}
\author[b,c]{Yufeng Li,}
\author[d,e]{Minglei Tong,}
\author[f]{Hongsheng Zhang}

\affiliation[a]{School of Physics and Technology, Wuhan University, Wuhan 430072, China}
\affiliation[b]{Key Laboratory for Computational Astrophysics,
National Astronomical Observatories, Chinese Academy of Sciences,
Beijing 100012, China}
\affiliation[c]{School of Astronomy and Space Science, University of Chinese Academy of Sciences, Beijing 100049, China}

\affiliation[d]{National Time Service Center, Chinese Academy of Sciences, Xian 710600}
\affiliation[e]{Key Laboratory of Time and Frequency Primary Standards, Chinese Academy of Sciences, Xian,710600}
\affiliation[f]{School of Physics and Technology, University of Jinan, 336 West Road of Nan Xinzhuang, Jinan, Shandong 250022, China}

\emailAdd{zhanghg.umd@protonmail.com}
\emailAdd{xilong.fan@whu.edu.cn}
\emailAdd{liyufeng1113@163.com}
\emailAdd{mltong@ntsc.ac.cn}
\emailAdd{sps$\_$zhanghs@ujn.edu.cn}

\abstract{We study the stochastic gravitational waves from string gas cosmology. With the help of the Lambert W function, we derive the exact energy density spectrum of the stochastic gravitational waves in term of tensor-to-scalar. New feathers with the spectrum are found. First, the non-Hagedorn phase can be ruled out by the current B-mode polarization in the cosmic microwave background. Second, the exact spectrum from the Hagedorn phase with a logarithmic term is shown to be unique in the measurable frequency range. Third, which is the most important, we find the string length can be constrained to be lower than 7 $\sim$ orders of that Planck scale.}

\begin{document}
\maketitle
\flushbottom

\section{Introduction} 
String theory  (see \cite{string} for a review) is the most hopeful scheme for complete quantum gravity, which is supposed to be valid up to the Planck scale. String length is a fundamental parameter in string theory, which is to be determined by experiment  \cite{witten}. It has an inverse relation with the string coupling constant $g_s$, and the effective string length in physical world of four dimensions has a relation with the extra dimension volume in string theory. In this letter, we propose a possible way to determine the effective string length through experiment.

Our work in based on string gas cosmology (SGC)  \cite{sgc}, which is a model of the early universe with minimal and crucial string inputs: a gas of closed strings \cite{blocks} as building blocks coupled to the space; the string oscillatory with winding modes as freedom; and T-duality as symmetry. SGC model can be embedded into string theory. The topics of extra dimension \cite{extra}, moduli fitting \cite{moduli}, elements of open string and D-brane \cite{brane}, the ``trans-Planckian'' problem \cite{trans-1}, and the  ``swampland'' problem \cite{swampland} are discussed in related literature.  As explained in  \cite{sgc}, T-duality implies the energy spectrum of string states is invariant when winding and momentum quantum numbers are interchanged with $ (n,m) \rightarrow (m,n)$, and the consequence of T-duality resolves cosmological singularities immediately as the  temperature in SGC has a maximum, the Hagedorn temperature $T_H$ \cite{hagedorn}. 

SGC is a thermal dynamics system of strings. With assuming the matter is in thermal equilibrium, it predicts an almost scale-invariant spectrum of cosmological scalar perturbations in the Hagedorn phase, and thus it provides an alternative scenario to slow-roll inflation for the origins of the large-scale structure. In fact, the mechanism for the generation of the primordial perturbations is intrinsically stringy, as it predicts a spectrum with red tilt, unlike a large and phenomenological blue tilt in spectrum from particle thermodynamic fluctuations.

Since the first gravitational wave signal from a binary black hole merger on 2015 September 14th detected by LIGO \cite{ligo}, tens of BBH mergers and one binary neutron star (BNS) signal have been identified by the LIGO-VIRGO network (see \cite{ligo-virgo}). In spired of that, the detection of stochastic gravitational-wave background (SGWB) becomes a hot topic recently, though that has been discussed in a series of previous works (e.g. \cite{allen,2008PhLB..663...17F,2012PhRvD..85j4024W,2011ApJ...729...59Z,2011ApJ...739...86Z,2014PhRvD..89h4046R,2019arXiv191109745M,2018PhRvD..98d4020F,fan}). Current network of ground-based GWDs have set directional limits on the SGWB \cite{ligo03}.

SGWB is also predicted from SGC, and a rough scale invariant power spectrum with a blue tilt is shown in (\cite{bluetilt-1, bluetilt-2}).  Recently, probing quantum gravity effect through stochastic gravitational wave background in SGC is studied in \cite{calcagni}. We revisit this topic in this work. With the help of Lambert W function, an exact energy density spectrum of gravitational waves background with the parameter of scalar-to-tensor is derived, and new feathers of the spectrum are then found. 

\section{The SGWB from SGC} 

We begin with the ansatz with a space-time metric containing both the linear scalar metric fluctuations $\phi$ and the gravitational wave fluctuations $h_{ij}$:
\begin{equation}
\label{fun-space}
ds^2 = a^2(\eta)((1 + 2\phi(x, \eta))d\eta^2 - [(1-2\phi) \delta_{ij} + h_{ij} ] dx_{i}dx_{j} ) ~,
\end{equation}
where $\eta$ is conformal time, $a(\eta)$ is the scale factor describing the background cosmology. We have chosen a gauge in which the metric corresponding to the cosmological perturbations is diagonal, and assumed that there is no anisotropic stress.

The inhomogeneities in SGC are not vacuum fluctuations, but rather thermal ones. Both of the scalar and gravitational waves (tensor metric) fluctuations are determined by the Einstein constraint equations, in terms of the fluctuations of the energy-momentum tensor. Specifically the gravitational fluctuations can be expressed as (see \cite{bluetilt-1})
\begin{eqnarray}
\label{fun-1}
&\Delta_{h}(k)& =  <|h(k)^2|> = \frac{16 \pi^2 G^2}{k^{4}} \left<\delta T^{i}_{j}(k) \delta T^{i}_{j}(k)\right> ~, (i \ne j)
\end{eqnarray}
where $G$ is Newtons gravitational constant, and $h(k)$ is the amplitude of each of the two polarization modes of gravitational waves. As explained in \cite{bluetilt-2}, this is the gravitational fluctuations at present time generated from the usual theory of cosmological fluctuations.

The correlation of Eq. (\ref{fun-1}) is calculated in \cite{bluetilt-1} in a fixed-coordinate. Here we express that in co-moving coordinate in term of the frequency $f$ as 
\begin{eqnarray}
\label{fun-2}
&\Delta_{h}(f)& = 16 \pi^2 G^2 \times {T \over {l_s^3}} \times (1 - \theta)\times \ln^2{\left[\frac{(1 -\theta)}{4\pi^2 l_s^2 f^2/c^2}\right]} ~,
\end{eqnarray}
where c is the light speed. We have set $\theta=T/T_{H}$, where $T=T(f_*)$ is the temperature when the mode with the frequency $f_*$ exits the Hubble radius. The power spectrum is independent on $a(\eta)$ and thus it is scale-invariant. As explained in \cite{bluetilt-2}, the temperature $T$ is independent on $f$ to the first approximation, and thus in this case, $T$, $\theta$ and $l_s$ are free parameters. In hagedorn phase with temperature close to the Hagedorn value, the power spectrum reduces to
\begin{eqnarray}
\label{fun-3}
&\Delta_{h}(f)& =  (\frac{l_{pl}}{l_s})^{4} \times \theta \times (1 - \theta) \times \ln^2{\left[\frac{(1 -\theta)}{4\pi^2 l_s^2 f^2/c^2}\right]} ~,
\end{eqnarray}
where $l_s$ and $\theta$ are the free parameters, with $ 0 < \theta <1$. A suspected blue tilt of the tensor model is predicted in (\cite{bluetilt-1, bluetilt-2}) from the logarithmic term in Eq. (\ref{fun-3}), however, we will show that this blue tilt is prohibited by the current B-polarization observation.

The energy intensity spectrum can be characterized by the dimensionless quantity as (see \cite{neil}) 
\begin{eqnarray}
\label{fun-4}
&\Omega_{0}(f)& = \frac{1}{12} \frac{2\pi f}{H_{0}} \times \Delta_{h}  ~,
\end{eqnarray}
where the Hubble constant at present is $H_0=3.2\times 10^{-18} h_{100}/sec$, and we use $h_{100}=0.65$.

The tensor-to-scalar $r$ in both Hagedorn phase and non-Hagedorn phase in SCG is defined as (see \cite{trans-1})
\begin{eqnarray}
\label{lambert1}
r(f) &=& (1 -\theta)^2 \times \ln^2{\left[\frac{(1 -\theta)}{4\pi^2 l_s^2 f^2/c^2}\right]} .
\end{eqnarray}
With rewriting it as
\begin{eqnarray}
\label{lambert2}
\frac{r^{1/2}}{4\pi^2 l_s^2 f^2} &=& \frac{(1 -\theta)}{4\pi^2 l_s^2 f^2} \times \ln{\left[\frac{(1 -\theta)}{4\pi^2 l_s^2 f^2/c^2}\right]} ~,
\end{eqnarray}
we find the $\theta$ can be solved inversely, and given by the so-called  Lambert W function. The prototype Lambert W function is the inverse function of $x=y \cdot e^{y}$, and the solution is $y=W(x)$, however for Eq. (\ref{lambert2}), which is of the form $x=y\cdot \ln(y)$, the inverse function is given as $y=x/W(x)$. We write that as below immediately
\begin{equation}
\label{lambert3}
\theta(f) = 1- \frac{r^{1/2}}{W(\frac{r^{1/2}}{4\pi^2 l_s^2 f^2/c^2})} ~~.
\end{equation}
The prototype Lambert W function has two branches, $W_{0}$ and $W_{-1}$, but for Eq. (\ref{lambert2}), the function only has one branch $W_{0}$ sine the term in the left of Eq. (\ref{lambert3}) is always larger than 0. Now, we have derived the energy density $\Omega_0$ in term of $r$, in both Hagedorn phase  and non-Hagedorn phase. 

The tensor-to-scalar parameter r can be determined by the primordial CMB B-mode polarization experiment. In 2018, the Planck team announced the result, a tightest tensor-to-scalar ratio of $r \leq 0.06$ is given in \cite{b-mode}, combining with the BICEP2/Keck Array BK14 data.
Constraining the SGWB through CMB observation is studied in \cite{trans-1} and  \cite{calcagni},  however, as they ignored the logarithmic term in Eq. (\ref{lambert1}), the constraining is loose. We give our analysis below, with using the exact spectrum derived above. 

\section{Rule out the non-Hagedorn phase} 
\begin{figure}[hbt!]
\centering
\includegraphics[width=100mm]{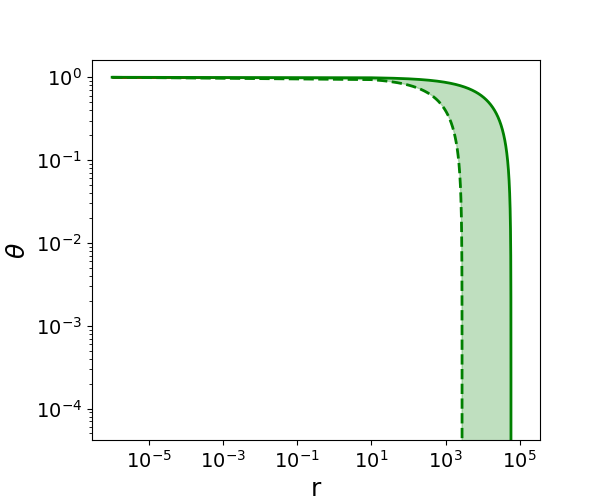}
\caption{\label{ratio}The dependence of $\theta$ with r, with the measurable frequency of $f \subset (10^{-10}Hz, 10^{10}Hz)$, and the string length $l_s \subset (l_{pl},l_{electron})$. The dash one represents the upper boundary while the solid line represents the lower boundary.}
\end{figure}

The non-Hagedorn phase is difficult to handle in SGC (e.g. see \cite{calcagni}), as the form of $T(k)$ is unknown. However, in this section, we show that the non-Hagedorn phase can be ruled out by the CMB data. It is reasonable to assume the string length $l_s \subset (l_{pl},l_{electron})$, where $l_{pl}$ is the Planck scale and $l_{electron}$ is the radius of electron, and with a measurable frequency $ f \subset (10^{-10}Hz, 10^{10}Hz)$,  $\theta$ from Eq. (\ref{lambert3}) can be illustrated in Fig. \ref{ratio}. Two observations: first, a steep drop, which is caused by the logarithmic term in (\ref{lambert1}), is observed within $r\subset (10^3,10^5$). However, the steep drop is in fact prohibited as the recent B-polarization experiment  \cite{b-mode} has given $r \leq 0.06$. Second, a constraint $\theta \geq 0.995$ is found in the region of $r \leq 0.06$. This implies only the mechanics of the mode exiting from the Hubble radius  from  Hagedorn phase with $T \geq 0.995 T_{H}$, is allowed, while the non-Hagedorn phase mechanics is ruled out.

\begin{figure}[hbt!]
\centering
\includegraphics[width=100mm]{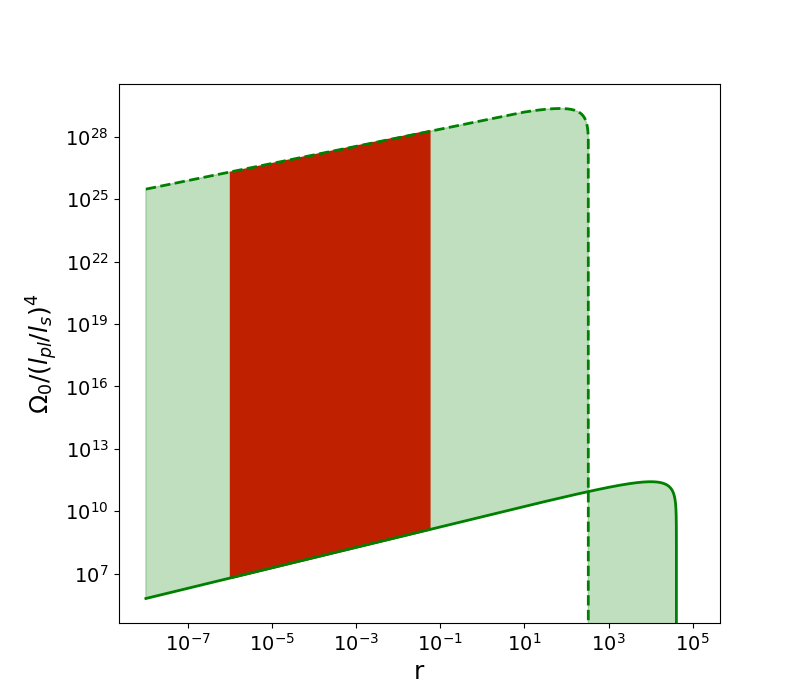}
\caption{\label{theta}The dependence of $\Omega_0/(l_{pl}/l_s)^4$ with r. We do that with a measurable frequency of $ f \subset (10^{-10}Hz, 10^{10}Hz)$, and a string length of $l_s \subset (l_{pl},l_{electron})$. The dash one represents the upper boundary while the solid line represents the lower boundary. }
\end{figure}

As the measurable frequency and the string length assuming above are quite loose, our conclusion is quite general. With the same assumption, we find a similar steep drop in the spectrum of $\Omega_0/(l_{pl}/l_s)^4$ in dependent of $r$ in Fig. \ref{theta}, which is obviously prohibited by CMB observations with the same reason. Thus, only the red region is the figure is left, from the aspect of CMB. \\

\section{SGWB detectability  from the Hagedorn phase} 

The spectrum of gravitational waves from SCG can be simply written in the form of $\Omega \sim f\times ln^2[(1-\theta)/f^2]$. It is clear that such a spectrum with a special logarithmic term is unique. However, we don't know how significant the logarithmic term makes in the  measurable frequency range ($10^{-4}Hz$, $10^{4}Hz$) of the current detector like aLIGO, and the upcoming detectors such as ET, DECIGO and LISA.

\begin{figure}[hbt!]
\centering
\includegraphics[width=120mm]{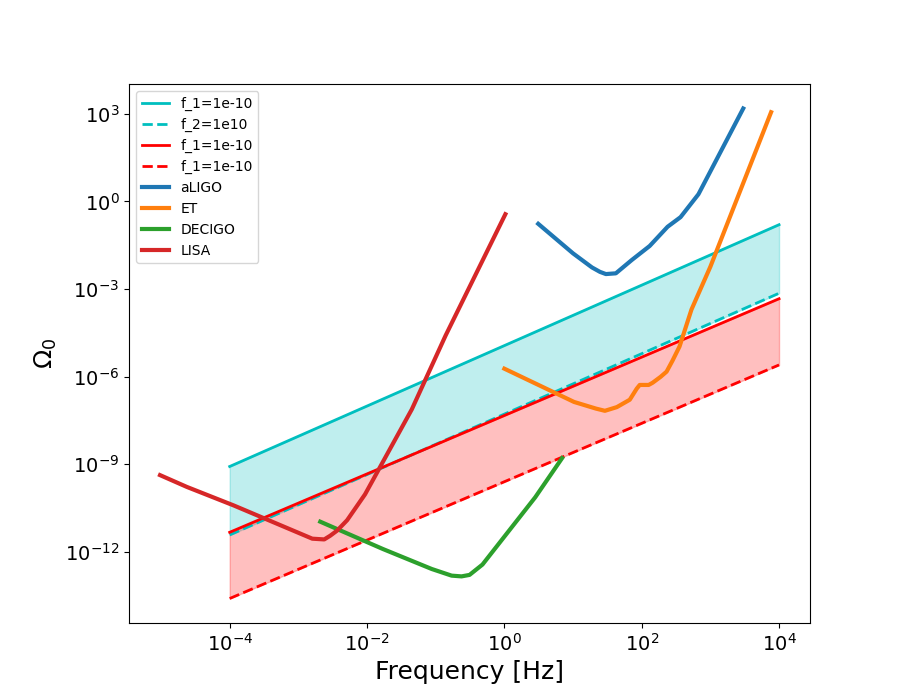}
\caption{\label{spectrum}The spectrum region of the energy density $\Omega_{0}$ of gravitational waves at present time from SGC with setting $l_s=10^{6} l_{pl}$ and $r \subset (10^{-6},0.06)$. The cyan region is the the case with the logarithmic term in Eq.(\ref{lambert1}), while the red region is the case without. The dash one represents the upper boundary while the solid line represents the lower boundary.  }
\end{figure}

The simplest way to see the uniqueness is comparing the spectrum with the one without such a term, which is of the power-law form $\Omega \sim f$. This is illustrated in Fig. \ref{spectrum}, with setting $l_s=10^{6} l_{pl}$ and $r \subset (10^{-6},0.06)$. The cyan region is the case with the logarithmic term, while the red region is the case without. It is clear that the spectrum region with the logarithmic term is higher than the one without, around 3 order. 

Thus, we see that the logarithmic term makes significant change in the spectrum in the measurable range. Thus, is fair to say that, in case such a spectrum is found, the origin of gravitational waves from SCG is confirmed.

\section{The string length constraining} 

The relation $\Omega_{0} \sim (l_{pl}/{l_s})^{4}$ in Eq. (\ref{fun-4}) implies the energy density spectrum is sensitive to the string length, and the smaller the string length, the easier the gravitational waves to be detected.  The detectability of GWDs for SGWBs radiation \cite{allen} could be quantitatively characterized by the ratio of `signal'~(S) to `noise'~(N) which is given by an integral over frequency $f$ after correlating signals for duration $T$:
\begin{equation}
\label{SN1}
\left( \frac { S } { N } \right) ^ { 2 } = \frac { 9 H _ { 0 } ^ { 4 } } { 50 \pi ^ { 4 } } T \int _ { 0 } ^ { \infty } d f \frac { \gamma ^ { 2 } ( f ) \Omega _ { \mathrm { 0 } } ^ { 2 } ( f ) } { f ^ { 6 } P _ { 1 } ( f ) P _ { 2 } ( f ) } ~,
\end{equation}
where the Hubble constant $H _ { 0 } =  3.2 \times 10 ^ { - 18 }  h  _ { 100 } / \rm sec $ is the rate at which our universe is currently expanding and $h_{100}$ is a dimensionless parameter for Hubble constant and is assumed to be 0.65 in this paper. $P_i(f)$ is the one-side noise power spectral density which describes the instrument noise for GWDs in the frequency domain.  $\gamma(f)$ denotes the overlap reduction function between two GWDs which encodes the relative positions and orientations of a pair of GWDs \cite{flanagan}.

Overall, we can infer from Eq.~\eqref{SN1} that the detectability of a network of GWDs to the SGWs are determined by the noise power spectral density together with the overlap reduction function of GWDs.  The detailed information and description of overlap reduction function of multiple detector pairs together with their sensitivity curves we used in later analysis could be found in \cite{fan}. In order to detect SGWBs with $5\%$ false alarm and $95\%$ detection rate, the total signal-to-noise ratio (S/N) threshold $S/{N_\mathrm{opt}}$ in Eq.~\eqref{SN1} should be 3.29 \cite{allen}.

\begin{figure}[hbt!]
\centering
\includegraphics[width=120mm]{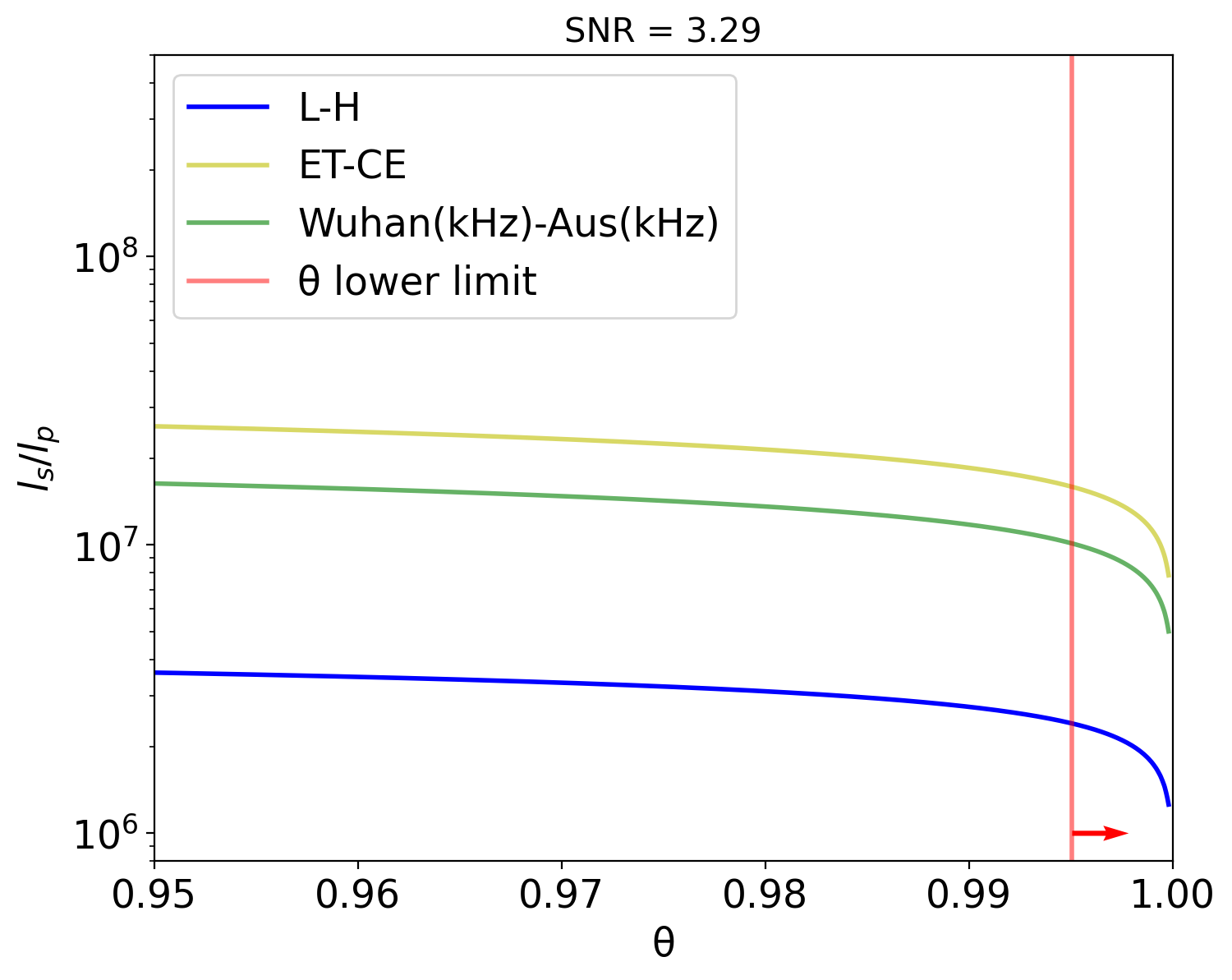}
\caption{\label{constrain}The string length $l_s$ versus the tensor-to-scalar $r$ given that network S/N of 3.29.  `L' is for `LIGO Livingston', `H' is `LIGO Hanford'. 'ET' represents the third generation detector Einstein Telescope, while 'CE' is Cosmic Explorer. Wuhan(kHz) means a kilo-Hertz detector located in Wuhan of China. Aus(kHz) represents the same kilo-Hertz detector located in Australia. The red vertical line represents the $l_s/l_{pl}$ at position of $\theta$ with $r=0.06$. }
\end{figure}

Given a fixed SNR threshold of 3.29, we could derive the relation between $l_s$ and $\theta$ in Eq (\ref{fun-2}) by Eq (\ref{fun-3})  as presented in Fig.\ref{constrain} for GWD networks by correlating their signals for 3 months $(T = 10^7 s \sim $3 months). By illustration, we simulated three GWD networks. They are respectively, the second generation GWD network, L(LIGO livingston)-H(LIGO Hanford); the third generation GWD network, ET(Einstein Telescope)- CE(Cosmic Explorer), and Wuhan(kHz)-Aus(kHz), which assumed 2 detectors with kilo-hertz detector sensitivity respectively located in Wuhan of China and Australia.

The relation $\Omega_{0} \sim \theta \times (1 - \theta)\times \ln^2{[(1 -\theta)]}$ gives the dependence of the sensitivity and $\theta$. As shown in the figure, the constraining capability for $\theta$ is almost independent on the sensitivity of GWDs. To be specific, lower right region, namely larger theta and smaller $l_s$, is easier to be tested as sensitivity of second generation detector network is enough to detect with higher SNR. On the other hand, upper left region, which corresponds to smaller theta and larger $l_s$, is harder to be explored, because GWDs with extremely high sensitivity are also limited to detect this region.

The relation $\Omega_{0} \sim (l_{pl}/{l_s})^{4}$ in Eq. (\ref{fun-2}) implies the energy density is sensitive to the string length, and the smaller the string length, the easier the gravitational waves to be detected. This could be seen from this figure that given the same $\theta$, the third generation detector network with higher sensitivity could detect larger string length $l_s$ with the same network SNR compared to the seconds generation detector network, which means detector networks with higher sensitivity has the ability to explore larger parameter space. And the constraining capability for $l_s$ is mainly determined by the GWD sensitivity.

With $r<0.06$, we find from Fig.\ref{constrain} that the string length $l_s$ is confined to be lower than $8\sim$ orders of the Planck scale from the third generation of GWD, while from the second generation of GWD, it is confined to be lower than $7\sim$ orders of the Planck scale. It is estimated in \cite{witten} that the string length might be of $15 \sim$ order of the Planck scale. However, if that is right,  finding the gravitational waves from SCG through the current the upcoming detectors is completely impossible.

\section{Conclusion} 
In conclusion, SGC is a thermal dynamics system of strings, and its prediction can be used to test string theory. The scalar perturbation from SCG remains it as one of the origins of large-scale structure. 

SGWB from SCG is studied in this work. With using the Lambert W function, we derived the exact energy density spectrum of SGWB in terms of the measurable tensor-to-scalar. With that spectrum, we found the non-Hagedorn phase in SCG is ruled out by the current B-polarization experiment data . As the logarithmic term plays a critical role, the spectrum from Hagedorn phase  was found to be unique  in the measurable frequency range. The most important, we found the string length can be confined to be lower than $7\sim $ orders of the Planck scale with the current B-polarization experiment data. \\

\acknowledgments
X. F. thanks R. Brandenberger for pointing out the references on  string gas cosmology scenario. X. F. is supported by the National Natural Science Foundation of China(under Grants No.11922303) and Hubei province Natural Science Fund for the Distinguished Young Scholars (2019CFA052).

\end{document}